\documentclass[12pt]{article}

\usepackage{times}
\usepackage{graphicx}
\usepackage[superscript,biblabel]{cite}
\usepackage{placeins}
\usepackage{lineno}
\usepackage{floatflt}
\usepackage[normalem]{ulem}

\begin{document}




\title{Autonomous Extraction of Millimeter-scale Deformation in InSAR Time Series Using Deep Learning}
\author
{Bertrand Rouet-Leduc$^{1,\ast}$, Romain Jolivet$^{2,3}$,\\ Manon Dalaison$^{2}$, Paul A. Johnson$^{1}$, Claudia Hulbert$^{2}$\\ \\
\normalsize{$^{1}$Los Alamos National Laboratory, Geophysics Group, Los Alamos, New Mexico, USA}\\
\normalsize{$^{2}$ Laboratoire de G\'eologie, D\'epartement de G\'eosciences, \'Ecole normale sup\'erieure}, \\
\normalsize{PSL University, CNRS UMR 8538, Paris, France}\\
\normalsize{$^{3}$Institut Universitaire de France, 1 rue Descartes, 75005 Paris.}\\
\normalsize{$^\ast$To whom correspondence should be addressed; E-mail:  bertrandrl@lanl.gov.}
}

\date{}
\maketitle

\newpage
\begin{abstract}


Systematic characterization of slip behaviours on active faults is key to unraveling the physics of tectonic faulting and the interplay between slow and fast earthquakes. Interferometric Synthetic Aperture Radar (InSAR), by enabling measurement of ground deformation at a global scale every few days, may hold the key to those interactions. 
However, atmospheric propagation delays often exceed ground deformation of interest despite state-of-the art processing, and thus InSAR analysis requires expert interpretation and \textit{a priori} knowledge of fault systems, precluding global investigations of deformation dynamics. 
Here we show that a deep auto-encoder architecture tailored to untangle ground deformation from noise in InSAR time series autonomously extracts deformation signals, without prior knowledge of a fault's location or slip behaviour.
Applied to InSAR data over the North Anatolian Fault, our method reaches  2 mm detection, revealing a slow earthquake twice as extensive as previously recognized.
We further explore the generalization of our approach to inflation/deflation-induced deformation, applying the same methodology to the geothermal field of Coso, California. 





\end{abstract}

\newpage

\section{Introduction}

 Faults slip in a variety of modes, from dynamic earthquakes to transient slow earthquakes and aseismic creep \cite{Avouac:2015bs,Burgmann:2018fg}. The classical picture of faults being either locked and prone to dynamic and damaging earthquakes, or unlocked and quietly creeping to accommodate tectonic stress, is evolving. Growing evidence indicates complex fault behaviours and interactions among and between modes of slip \cite{Jolivet:2020ku}. Evidence includes fault segments hosting both slow and dynamic earthquakes, as well as slow earthquake preceding and possibly triggering the nucleation phase of dynamic earthquakes \cite{McLaskey2019a,Obara2016,scholz_2019}.  Answering a number of fundamental questions such as what controls the slip mode on a fault, whether there exists a continuous spectrum of slip modes on faults, and what determines the possible evolution of a slow earthquake into a dynamic seismic rupture, requires exhaustive characterization of all slip phenomena. Interferometric Synthetic Aperture Radar (InSAR) holds the promise of continuous geodetic monitoring of fault systems at a global scale, which may well hold the key to address these questions. However, although the data exists, current algorithms are not suited for global monitoring because they require time-consuming manual intervention, and the final product requires exhaustive expert interpretation. 

InSAR is routinely used to measure ground deformation due to hydrologic, volcanic, and tectonic processes\cite{Fialko2012,Chaussard2014,Jolivet2015a}. The apparent range change in the satellite Line-Of-Sight (LOS) between two SAR acquisitions is, after corrections from orbital configurations and topography, the combination of atmospheric phase delay and ground deformation. Rapid, large-amplitude deformation signals such as coseismic displacement fields often exceed the amplitude of other known sources of noise\cite{Simons2002}. Similarly, slow but steady accumulation of deformation over long periods of time may be quantified using InSAR either through stacking, e.g. \cite{Peltzer2001}, or time series analysis\cite{Jolivet2012,Weiss:2020aa}. However, detecting low-amplitude deformation related to transient sources such as slow slip events, episodes of volcanic activity or hydrologic related motion remains challenging and requires significant human intervention and interpretation \cite{Chaussard2014,Rousset2016,Khoshmanesh2018}. Measuring Earth surface deformation is fundamental to characterizing diverse tectonic processes and the impact, as well as surface and undergound changes induced by human activities. 

The most pressing issue in InSAR processing for small, mm-scale, deformation monitoring remains the separation between atmospheric propagation delays and ground deformation. Spatial and temporal variations in atmospheric pressure, temperature and relative humidity modify the refraction index of the air, resulting in spatial and temporal delay variations in the two way travel time of the radar carrier between a SAR imaging satellite and the ground \cite{Hanssen2001,Doin2009a}. Such delays directly affect the phase of an interferogram, that combines two SAR acquisitions. Atmospheric propagation delays in a single interferogram can be equivalent to tens of centimeters in apparent range change \cite{Hanssen2001}. Current correction methods based either on empirical estimations, e.g. \cite{Elliott2008,Cavalie2008} or on independent data, e.g. \cite{Li:2006ia,Jolivet2011,Yu2018,Shen:2019aa} reduce the contribution of the stratified atmosphere -- the long wavelength atmospheric perturbation that, to first order, correlates with topography. Nonetheless, remaining delays, corresponding to the turbulent portion of the troposphere may represent centimeters of apparent range change. Propagation delays in the atmosphere decorrelate after periods of 6 to 24 hours, as shown by the temporal structure function of GNSS zenith delays \cite{Emardson2003}. Therefore, remaining tropospheric delays, which are coherent in space, can be considered random in time given the time span between SAR acquisitions (e.g. 6 days for Sentinel 1, 46 days for ALOS-2). However, it can be shown that, because of potential temporal aliasing \cite{Doin2009a} and loss of spatial coherence of the radar phase echo, spatio-temporal filtering can lead to biased results. Therefore, deciphering a consistent, days- to month-long transient signal in time series of InSAR remains a critical challenge, especially when automation is envisioned. 

Here we describe a deep learning-based method to automatically detect and extract transient ground deformation signals from noisy InSAR time series. Our approach, based on a purely convolutional auto-encoder, is specifically designed for removing noise in InSAR time series. In the following, we consider the evolution of the interferometric phase with time with respect to a reference both in space and time. We consider classical Small Baseline (SBAS)-like approaches for the reconstruction of the time series e.g.\cite{Berardino2002,Dalaison:2020aa}. Approaches include inversion from a sequence of SAR interferograms previously corrected from orbital and topographic contributions \cite{Simons2007}, with a first-order atmospheric correction derived from global atmospheric re-analysis products \cite{Jolivet2011,Jolivet2014}.

Convolutional neural networks (CNN) are central to most recent dramatic advances in computer vision and natural language processing. Auto-encoders have been developed to create sparse representations of data --the model copies its input to its output through a bottleneck that forces a reduction of dimension equivalent to a compressed knowledge representation of the original input, enabling noise removal. Of note are recent developments applied to classify InSAR data in order to detect ground uplift and subsidence, and specifically to identify volcanic unrest \cite{Schwegmann2017,Anantrasirichai2018,Anantrasirichai:2020aa}. Although promising, these developments do not make use of the different temporal signatures of signals of interest to reconstruct de-noised deformation patterns. Our auto-encoder takes as input a noisy InSAR time series reconstructed from successive SAR acquisitions, and outputs accumulated ground deformation taking place during the time series interval, with the atmospheric noise removed. 
 
In the following, we first introduce the notion of auto-encoders before describing the architecture of our neural network. We then describe our training set and perform preliminary tests on synthetic data. We finally highlight the efficiency of our algorithm on two reconstructed time series of ground deformation, the first one derived from COSMO-SkyMed acquisitions and the second one derived from Sentinel 1A-B SAR acquisitions. 

\section{Results}

\subsection{Description and Validation of the deep auto-encoder}

\subsubsection*{Auto-encoder architecture}

Our goal is to extract ground deformation from noisy InSAR time series. We assume that input time series are the combination of three physical contributions: ground deformation, the stratified component of the atmosphere and the turbulent component of the atmosphere. In most cases, the stratified component can be corrected for using Global Atmospheric Models (hereafter referred to as GAMs, often corresponding to re-analysis products), e.g. \cite{Jolivet2014}, or Global Navigation Satellite System (GNSS) data, e.g. \cite{Onn2006}, for instance. However, such a correction is often incomplete and part of the remaining, often turbulent, atmospheric delays may still correlate with topography. Attempts have been made to estimate tropospheric delays using multi-spectral radiometric data \cite{Li:2006ia}; however, the acquisition of such independent data must be coincident with the SAR acquisition and over a terrain with minimal cloud cover for optimal performance, conditions rarely met. In addition, it can be shown that GAMs-derived correction sometimes worsen the situation as the local estimate of the state of atmospheric variables may be incorrect \cite{Jolivet2014}. 

\begin{figure*}[!ht]
    \begin{center}
        \includegraphics[width=0.9\linewidth,trim= 0 0 0 0]{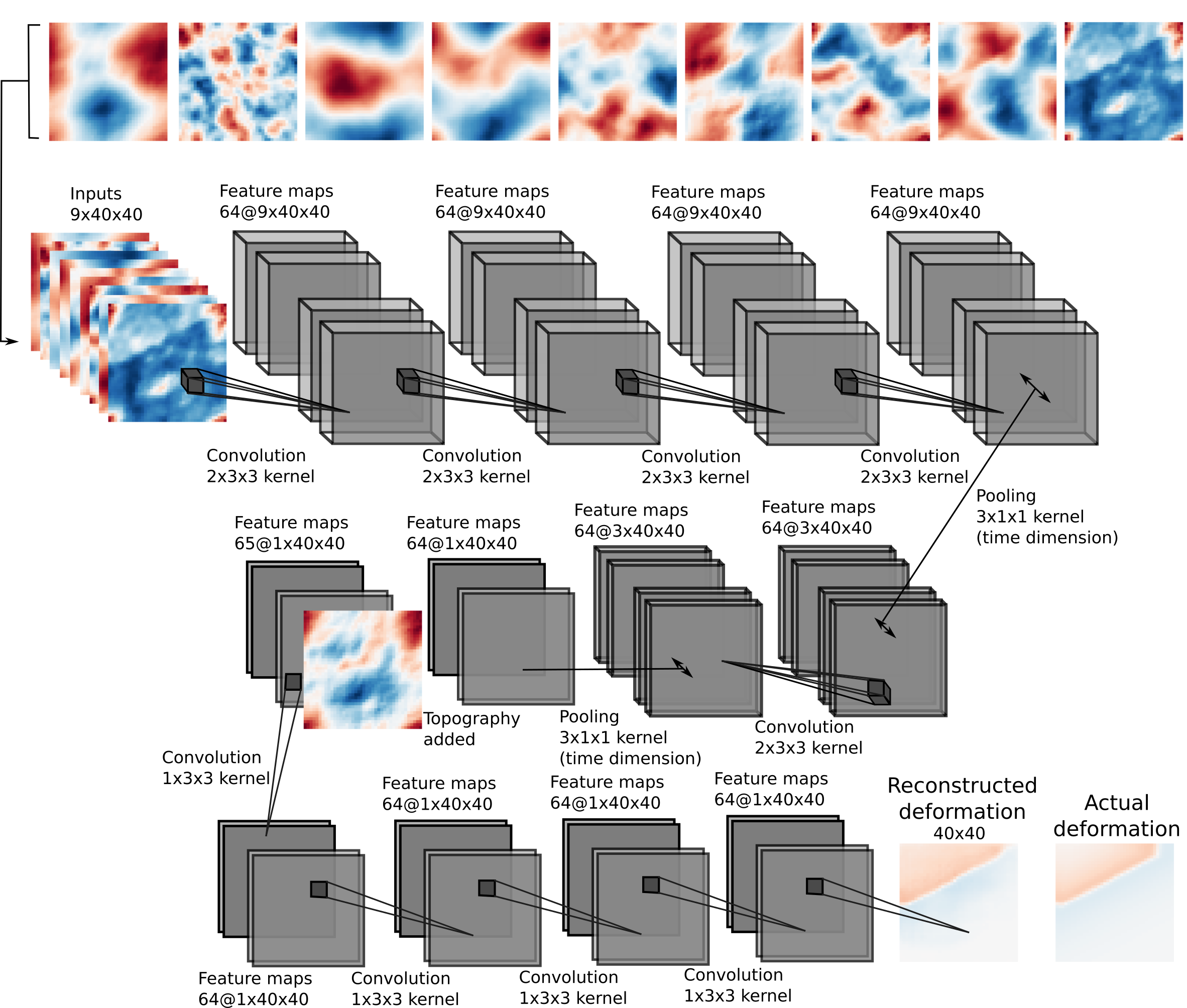}
        \caption{\footnotesize{\textbf{Auto-encoding InSAR time series.} Schematic of our deep learning model. Top row (left to right): a sequence of synthetic InSAR time series on which the model is trained, where ground deformation signal is corrupted with atmospheric noise, including turbulence and layering of the atmosphere. Second to fourth rows: architecture of our model. Our model is purely convolutional with progressive pooling on the time dimension during the encoding. After time is removed, at the 7th layer, ground elevation is added as a secondary input. Fourth row: The last layers of the model are tasked with decoding ground deformation accumulated during the input time series, here compared with actual deformation that takes place in the synthetic time series shown above. A  detailed description of this neural network can be found in the methods section.}}
        \label{fig1}
    \end{center}
\end{figure*}

Our deep learning model must recognize transient deformation in InSAR time series in the presence of remaining atmospheric noise. To this end, it must distinguish the spatial and temporal statistical differences between deformation signals and atmospheric patterns. As mentioned above, the structure of atmospheric delays decorrelates for periods longer than 6 hours\cite{Emardson2003}. Therefore, as ground deformation related to transient tectonic events might take place over seconds to minutes for dynamic rupture, to weeks or months or even years for slow slip events \cite{Jolivet2013,Rousset2016,Khoshmanesh2018,Materna:2019kk}, and remains until further ground deformation occurs, the temporal signature is very different from that of atmospheric delays. We make use of this different temporal signature to learn appropriate filters to remove atmospheric perturbations and extract ground deformation in InSAR time series. 


Here, we build and train an auto-encoding architecture to directly output the deformation signal, formulating the problem as a regression task. We rely on the following assumptions: (1) atmospheric delays are random in time, considering two successive SAR acquisitions, (2) ground deformation has a temporal coherence considering the rate at which SAR images are acquired and (3) part of the atmospheric delay correlates with topography. We therefore use as inputs a time series of interferometric phase change and a map of ground elevation to produce a time series of cumulative surface displacements.

In order to separate deformation from atmospheric delays, we developed the deep learning architecture shown in Fig. 1. This architecture consists of 11 purely convolutional layers. The first 6 layers of the model are tasked with encoding signals that are persistent in time, by progressively removing the time dimension of the input. The remaining layers decode the ground deformation map. In short, we build a model tasked with reconstructing ground deformation given input InSAR time series and ground elevation from noisy input.

Initially developed for feature extraction by projecting high-dimensional datasets onto a lower-dimension manifold by forcing the reconstruction of the data through a bottleneck in deep learning architectures \cite{Vincent2008}, auto-encoders have also evolved into powerful denoising \cite{Feng2014,Zhang2017} and image enhancing techniques \cite{Cui2014,Mao2016}. In this work we exploit this aspect of deep learning auto-encoding and tailor it to the problem of cleaning InSAR time series, building a deep learning auto-encoder to effectively automate the design of filters in time and space to recover ground deformation. 

\subsubsection*{Training on synthetic data}

Because deep learning models require large amounts of data and there exists no ground truth for InSAR time series, we rely on synthetic data to train the deep auto-encoder. The synthetic data are randomly-generated cumulative surface deformation time series mimicking 9 successive `acquisitions'. These cumulative deformation maps include surface displacements in the line of sight (LOS) due either to a slipping fault with random latitude and longitude (position in a virtual box), depth, strike angle, dip angle and width (based on Okada's model \cite{Okada1992}) or to an inflating or deflating point source (Mogi's model \cite{Mogi1958}). Deformation onset occurs at a random time with a random duration within the time series, excluding the first and last acquisitions which are taken as non-deforming reference by the model. The model is therefore tasked with finding cumulative deformation in the 7 middle acquisitions of the time series arising from a wide variety of transient processes. We then corrupt each map of these ground deformation time series with different noise signals. At each time step we create both turbulent and stratified synthetic atmospheric delays. Spatially correlated Gaussian noise mimics delays from atmospheric turbulence of various length scales \cite{lohman2005,Sudhaus:2009jw} (Fig. 1 top row) and a quadratic function of pixels' elevation mimics the atmospheric delays that correlate with topography \cite{Hanssen2001,Bekaert:2015eta}. Each of the steps of the time series results from a random realization of noise built following these assumptions. 

We train two independent models with the synthetic time series of deformation, one tasked with recovering point source deformation, the other with recovering deformation on faults. All other phase delays are identified as noise. Both models are trained to map synthetic noisy time series to the synthetic cumulative displacements.  We trained our deep auto-encoder with $2.5 \times 10^7$ randomly generated time series for which we provide as input the apparent line of sight (LOS) deformation time series, corrupted by the sum of synthetic noise described above. The training includes a LOS with random orientation (30-45 degrees incidence and any azimuth), so that the model is directly trained for various SAR satellite configurations and for any fault azimuth. The output is the target ground deformation accumulated during the time series. All 482,185 trainable parameters are adjusted during that training phase with the Adam variation of stochastic gradient descent \cite{Kingma2014}.

We note that our deep auto-encoder only considers time series of 9 acquisitions, as a good compromise between the tractability of the training phase (the time series are long enough for the model to learn the temporal differences between signal and noise) and a minimum duration of the analysed time series of displacements. When working with longer time series of $n$ acquisitions, we apply the algorithm using a sliding window with a width of 9 time steps and obtain $n-8$ images of cumulative deformation. In this way, our model acts as a moving integral of actual deformation.

\subsubsection*{Performance on synthetic data set}

Once trained, we test the deep auto-encoder on synthetic realizations of time series that have not been used to train the model. We randomly generate 3200 time series of 9 time frames using the same procedure as that described for the training phase. For each of the 3200 time series, we evaluate the signal-to-noise ratio (hereafter referred to as SNR) as the ratio of the mean absolute ground deformation and the standard deviation of the noise. We then apply the deep auto-encoder to these time series in order to evaluate the performance of the model. We evaluate the resulting, cleaned time series using the coefficient of determination R$^2$, a standard regression metric, equal to 1 for a perfect reconstruction, 0 for a reconstruction no better than the empirical average, and negative for the worst reconstructions (for example reconstructions anti-correlated with the ground truth).

\begin{figure*}[!ht]
    \begin{center}
        \includegraphics[width=\linewidth,trim= 0 0 0 0]{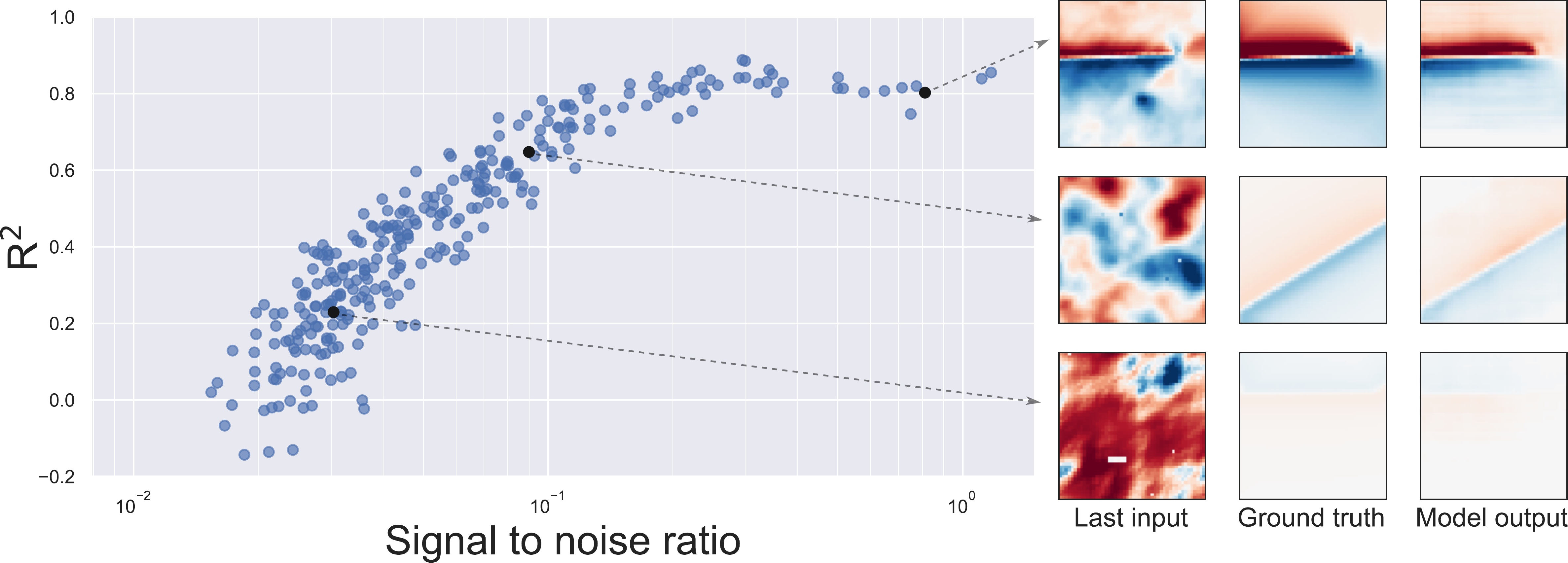}
        \caption{\footnotesize{\textbf{Performance on synthetic test data.} Left: Performance of the reconstruction of fault deformation by our deep auto-encoder, on synthetic noisy time series, as measured by the coefficient of determination R$^2$ (a metric of goodness of fit that compares model error with the error of predicting the empirical average) as a function of signal to noise ratio (SNR, see Methods). Each point represents the average over 64 synthetic time series with the same SNR. Right: examples of the data (last frame of the input), the ground truth and its reconstruction, for different signal to noise ratios. Note that the model outperforms the eye, recovering with reasonable fidelity (R$^2>0.6$) deformation signals with SNRs below 0.1.}}
        \label{fig2}
    \end{center}
\end{figure*}

We find that the deep auto-encoder applied to synthetic data accurately reconstructs deformation signals on faults, even in circumstances very challenging to expert interpretation (SNRs lower than 1; Fig. 2). For SNRs above 0.1, our algorithm provides a very accurate reconstruction ($0.6<$R$^2 <1.0$) of the cumulative ground deformation. For very low SNRs (0.1 and below), no signal can be visually observed, while the goodness of fit is still correct and the overall deformation signal is recovered down to SNRs of approximately 0.02, below which our model fails. Therefore, our architecture allows us to exceed the ability of the expert eye to detect signals in noisy time series of deformation, provided their noise structure resembles the training set.

In the following we show the application of our auto-encoder to two case studies that have been independently analyzed by InSAR experts. 

\subsection{Application to real data}

\subsubsection*{Extracting deformation from a slow earthquake along the North Anatolian Fault, Turkey}

Our deep auto-encoder is trained to isolate and reconstruct cumulative ground deformation signals in 40x40 pixel time series of 9 acquisitions. However, a fundamental property of purely convolutional deep learning models is that the filters they learn do not depend on input size. As a result, we can create an auto-encoder with exactly the same architecture as the model described in Fig. 1, but with an input size matching the number of pixels in the InSAR time series of interest. Because the parameters of the model do not depend on input size, we can copy every parameter (i.e. weights and biases of the filters) of the model trained on synthetics to the new model, which can then be applied to InSAR data of any size.

\begin{figure*}[!hb]
\vspace{-1cm}
\begin{center}
        \includegraphics[width=0.9\linewidth]{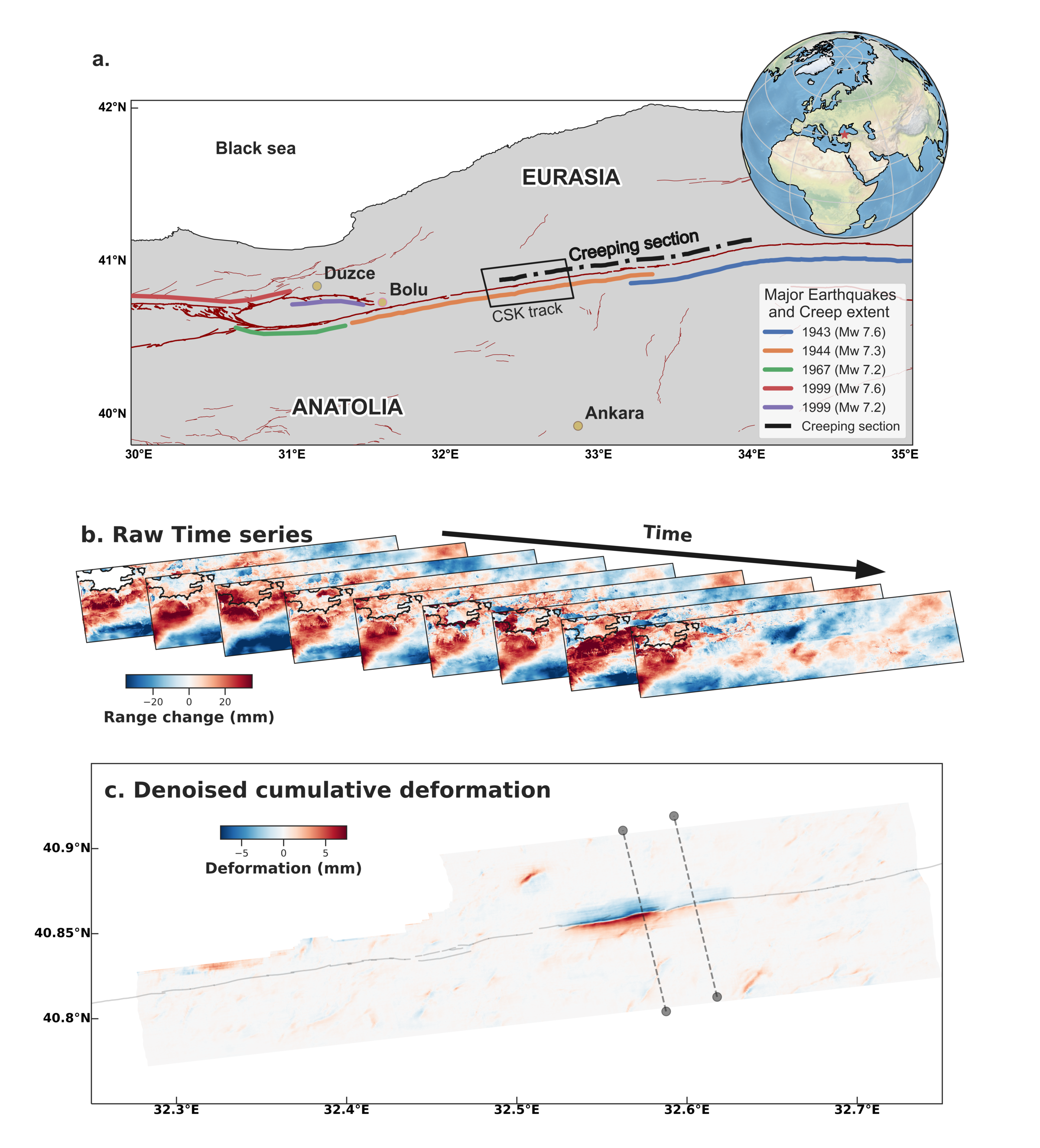}
        \caption{\footnotesize{\textbf{Application to real data: the North Anatolian Fault 2013 slow earthquake.} In order to identify ground deformation signals in the noisy COSMO-SkyMed InSAR time series, we create a deep auto-encoder that has an input size equal to the size of the acquisitions, 200x650 pixels, and the same parameters as the auto-encoder trained on synthetic data, shown in Fig. 1. Inputs are the InSAR time series. The auto-encoder outputs ground deformation (bottom plot). The ground deformation is manifest as an offset across the fault. The deep auto-encoder finds a strong slip signal of about 1 cm (in LOS) on the fault, in agreement with previous expert analysis of the time series \cite{Rousset2016}, with no \textit{a priori} knowledge of the fault's existence. {\bf a.} Seismic setting of the region of the creeping section of the North Anatolian Fault. Thick red lines are the main faults of the NAF system, separating the Eurasia plate from the Anatolia microplate. Thin red lines are other mapped structures. Colored lines indicate the extent of historical ruptures. {\bf b.} Input raw time series from COSMO-SkyMed data (subset of the data from Rousset et al. 2016). Color is the apparent range change between the ground and the satellite. {\bf c.} Denoised cumulative deformation as output by the deep-autoencoder. The color scale shows ground deformation in the direction of the LOS. Dark lines are the surface trace of the NAF, shown here for reference. Thin dashed lines indicate the cross-sections shown in figure \ref{fig4}.}}
        \label{fig3}
    \end{center}
\end{figure*}

Here, we apply the model to a time series built from images acquired by the COSMO-SkyMed constellation over the central section of the North Anatolian fault in Turkey (Fig. \ref{fig3}). This major plate boundary fault accommodates the motion of rotation of the Anatolia plate with respect to Eurasia and has ruptured in large, Mw 7 earthquakes multiple times over the last century \cite{Stein:1997du}. An 80-km-long section of the fault has been slipping aseismically, at least since the 1944, Mw 7.3, earthquake located near the small town of Ismetpasa \cite{Ambraseys1970}. {\it In situ} measurements based on creepmeters indicate that this fault experiences transient aseismic slip episodes \cite{Altay:1991wb,Bilham2016,Cetin:2014hp}. 

Rousset et al. produced an approximately one year long time series from COSMO-SkyMed SAR acquisitions and detected a significant slow slip episode that lasted one month during 2013 with a maximum of 2 cm of fault-parallel slip \cite{Rousset2016}. Average long-term velocity maps covering the whole region derived from InSAR data show aseismic slip over an 80-km-long section of the fault. This average relative displacement was found to result from successive transient events \cite{Rousset2016,Bilham2016}, which were not apparent in data from older constellations of SAR satellites due to the long time span between acquisitions. 

In the InSAR time series processed by Rousset et al., large atmospheric delays are apparent, despite careful correction of atmospheric delays using ECMWF re-analysis products \cite{Jolivet2011,Rousset2016}. Therefore, knowledge of the fault location was key in the interpretation of the surface displacement field.

Here we revisit the same time series in order to assess if our model is able to recover the known surface slip in real time series of data. We stress that we do not provide the location of the fault to the model.

With no human intervention and no \textit{a priori} knowledge of the local tectonics and fault location, the model automatically isolates and recovers clean deformation signals where expert analysis previously found signal attributed to tectonic activity (Fig. \ref{fig3}). Importantly, the recovered deformation is obtained after training only on synthetic data and with no further fine tuning on real data.  Our model finds up to 1.5 cm line of sight relative displacement across the fault, that we interpret as the signature of surface slip, as previously found \cite{Rousset2016}.

\FloatBarrier
\begin{figure*}[!ht]
    \begin{center} 
        \includegraphics[width=0.9\linewidth]{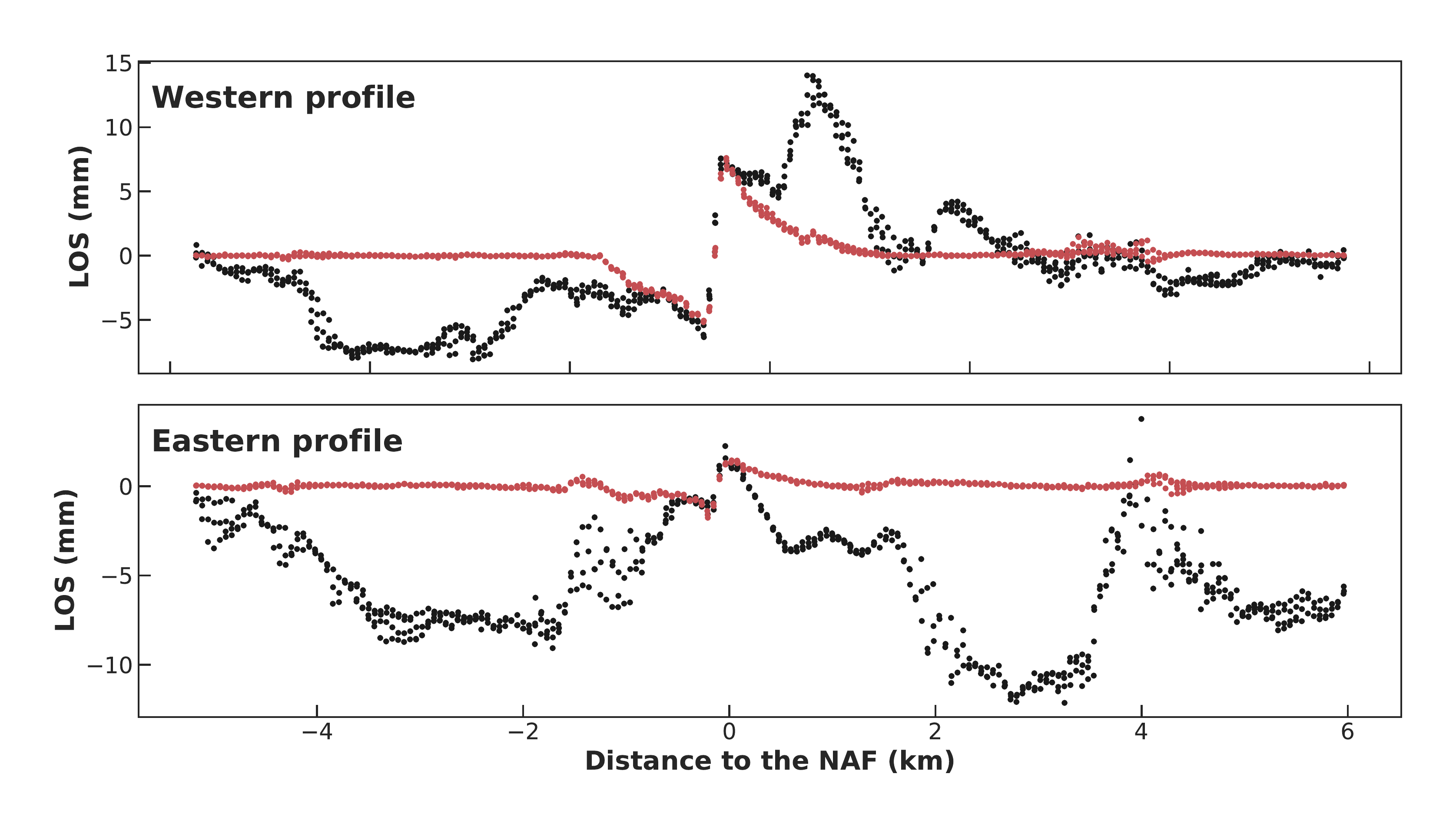}
        \caption{\footnotesize{\textbf{Application to real data: the North Anatolian Fault 2013 slow earthquake.} LOS deformation along fault-perpendicular cross sections. Location of the cross-sections are shown in Fig. \ref{fig3}. Black dots are the difference between phases averaged over acquisitions between September 5th and 21st, 2013 and over acquisitions between August 4th and 28th, 2013, taken along a fault perpendicular line. The main slow slip event detected by Rousset et al. occurred during this period. Red dots are the output of the model highlighting the cleaned deformation pattern. The sharp offset in the input InSAR data observed exactly on the fault was interpreted as a slow slip event by Rousset et al., in spite of the very high noise level presumably caused by atmospheric delays. Such interpretation was only made possible owing to knowledge of the location of the fault and knowledge that this segment of the North Anatolian Fault slips aseismically. Our model knows neither and automatically extracts actual ground deformation. Our current model interprets wavelengths longer than a kilometer as noise, although experts might interpret those as the signature of slip at depth.}}
        \label{fig4}
    \end{center}
\end{figure*}
\FloatBarrier

Fault perpendicular cross sections illustrate that even in regions where slip would not have been convincingly identified by an expert (Fig. \ref{fig4}), our model recovers 2 mm of slip, extending the previous estimate of along-strike length of this slip event. Rousset et al. identified a 10-km-long slow slip event while the deep learning model determines that the portion that slipped was 15-20 km in length. Interestingly, the new 2 mm slow slip we find is on a segment adjacent to the previously identified 1 cm slow slip, and the two segments are separated by a kink on the fault, suggesting an interplay between fault geometry and slip \cite{Jolivet2015,Romanet2020}. 

\subsubsection*{Extracting ground deformation signal at the Coso geothermal system, California}

In a second example, we use our deep learning architecture to detect surface deformation caused by underground pressure changes. As above, our model is trained on several million examples of synthetic noisy InSAR time series. In this case, surface deformation is modeled by a point pressure source using Mogi's equation of elastic deformation\cite{Mogi:1958wk}, corrupted as before by synthetic atmospheric delays. Mogi pressure sources are used extensively for the modeling of volcanic inflation and deflation signals, e.g. \cite{Segall:2010,Grandin:2010gl}. Further, the combination of multiple sources allows one to model complex subsidence/uplift patterns. 

After training exclusively on synthetic data, we apply our model to real data from the Coso geothermal field (California, USA), again without further training (details on the InSAR processing are in the Methods section). Because InSAR time series may be very noisy, even after correcting predicted atmospheric effects \cite{Jolivet2014}, analysis of inflation or subsidence of less than a few centimeters per year in InSAR have relied to date on deriving long term cumulative deformation \cite{Hoyt:2020aa}, such that random atmospheric delays cancel out. Detecting transient subsidence and uplift signals in SBAS time series below a few centimeter remains challenging, just as it does for faulting. 

\begin{figure*}[!ht]
    \vspace{-2.5cm}
    \begin{center}
        \thispagestyle{empty}
        \includegraphics[width=0.8\linewidth]{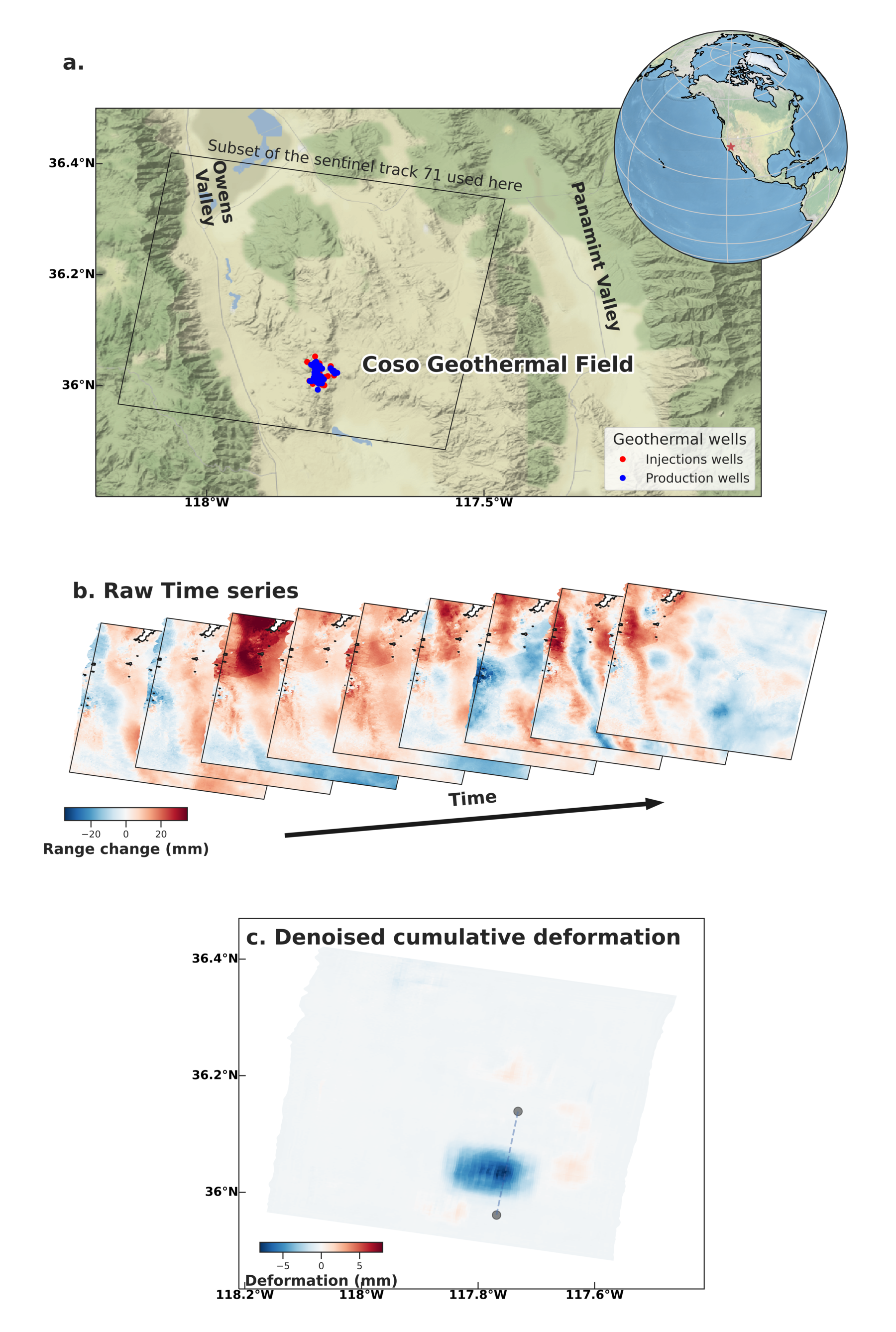}
        \caption{\footnotesize{\textbf{Application to real data: the Coso Geothermal Field in California.} After training our deep auto-encoder architecture exclusively on synthetic InSAR time series of point sources of deformation corrupted with atmospheric noise, we apply it to the time series obtained from Sentinel 1A-B from 2016-04-14 to 2016-11-16, that spans the Coso Geothermal Field in California. Our model detects a transient episode of subsidence of 5 to 7 mm (in line of sight), where the operational wells are located, with no \textit{a priori} knowledge of the area. {\bf a.} Geographic setting with the coverage of the subset of the Sentinel 1 track used here. Red and blue dots indicate the geothermal wells, respectively for injection and production. Background color is the terrain rendering from Stamen ({http://maps.stamen.com}). {\bf b.} Input raw time series of 9 successive images from Sentinel 1 data. Color is the apparent range change between the satellite and the ground along the LOS. {\bf c.} Denoised cumulative deformation as output by our deep auto-encoder. Color is ground deformation in the LOS. The thin dashed line indicates the location of the cross section shown in the Supplementary.}}
        \label{fig5}
    \end{center}
\end{figure*}

As with identifying deformation on faults, our model is able to disentangle actual ground deformation from atmospheric noise at short time scales, with a resolution of a few millimeters. In Fig. \ref{fig5} we show the application of our deep denoising model to a time series over Coso in 2016. Contrary to what could be inferred from long term cumulative deformation, we find that ground subsidence at Coso is primarily due to transient episodes of deformation.  The cumulative deformation from these transients we detect account for most of the cumulative deformation observed in the data (see Supplementary for details and for other examples of transient deformation). Interestingly, we find a number of transient events that are constituted of well separated pressure sources, in agreement with geochemical observations showing that the geothermal field is constituted of isolated reservoirs\cite{adams2000geologic}.

\FloatBarrier

\section{Discussion}

As the properties of the atmosphere cannot be measured at the same spatial and temporal resolution as InSAR acquisitions, InSAR time series still contain large amplitude atmospheric delays, on the order of centimeters, in spite of recent marked improvements in atmospheric correction and processing strategies \cite{Jolivet2014,Shen:2019aa}. For this reason, expert processing and analysis is required to interpret InSAR data. Furthermore, since the onset of the Sentinel 1 mission, the amount of available InSAR data has grown at a pace that is already challenging the ability of the community to process and analyze it, and the upcoming NISAR mission will increase the amount of available InSAR data several fold. Therefore, significant effort has been put into developing strategies to build time series with such vast data sets, e.g. \cite{Weiss:2020aa,Flatsim,Dalaison:2020aa}. Nonetheless automatic, autonomous InSAR interpretation methods are poised to become essential, if just to leverage the increasing spatial and temporal resolution of the data.

We note that several avenues of improvement should improve the ability of our neural network to detect finer and finer deformation signals in the future. First, we did not include sources of noise representative of ionospheric perturbations. The total electronic content of the ionosphere introduces a differential delay in interferograms that can bias analysis further \cite{Hanssen2001}. Although this effect is more pronounced for L-band SAR satellites \cite{Hanssen2001,Shen:2009ds}, long-wavelength ionosphere delays can be problematic for large images acquired with C-band SAR systems such as Sentinel 1 \cite{Liang:2019hv}. Although these delays can be corrected for using techniques such as the range split-spectrum method \cite{Fattahi:2017fp,Liang:2019hv}, the structure of the remaining noise associated with imperfect corrections must still be evaluated and could then be used in the training of our model. Second, we considered atmospheric turbulence to be isotropic and equivalent everywhere in the image (i.e. noise is second-order stationary) while some anisotropy can be observed in the phase delay of some interferograms. However, such anisotropy depends on the scale of the image observed, which would involve complex considerations in the construction of an adequate tropospheric noise model to train our model. In general, any improvement in the forward modeling of the nature of noise in InSAR should lead to a significant improvement in the detection capability of the models. 


Finally, the receptive field of the autoencoder and the pixel size of the input InSAR data restrict the size of the deformation signal that can be deciphered. For instance, interseismic deformation related to loading of a fault by plate motion extends over 10s of kilometers, e.g. \cite{Peltzer2001,fialko2006,Jolivet2012}. Additional developments may be necessary for the detection and cleaning of long wavelength deformation patterns.

The initial application of our method on InSAR time series enables the direct observation of a slow earthquake, refining previous estimates, autonomously and without prior knowledge. In particular, we expect that the ability to systematically observe fault and pressure source deformation at a global scale will further the understanding of hydrologic, volcanic and tectonic processes, and may bring us closer to bridging the observational that exists for transient surface deformation.

\bibliographystyle{naturemag}
\newpage
\bibliography{main}

\subsection*{Acknowledgements}
B. R.L.'s work was funded by Institutional Support (LDRD) at Los Alamos (20200278ER). R. J., M. D., C. H. were supported by the European Research Council (ERC) under the European Union’s Horizon 2020 research and innovation program (Geo-4D project, grant agreement 758210). C.H. was also supported by the CEA-ENS Yves Rocard LRC (France). P. J. was supported by DOE Office of Science (Geoscience Program, grant 89233218CNA000001).

\subsection*{Author contribution}
Author order uses the remote sensing convention of author contribution. B. R.L. and R.J. formulated the problem as a deep denoising task. B. R.L. created the deep learning model and applied it on real InSAR data, with help from R. J., M. D. and C. H.; R. J. implemented the synthetic data used for training the model and processed the COSMO SkyMED InSAR data; M. D. processed the Sentinel 1A-B data. All the authors analyzed the results and wrote the paper.

\subsection*{Data availability}
All the InSAR data used here is freely available from the European Space Agency. The COSMO-SkyMED archives and the Sentinel 1 data can be found at https://earth.esa.int.

\subsection*{Code availability}
The synthetic data used to train the model are based on the open source code CSI from R. Jolivet and can be found at http://www.geologie.ens.fr/~jolivet/csi/
The deep learning model has been developed using the open source Python package Tensorflow.
\newpage

\end{document}